\documentstyle[12pt,russian]{article}
\begin {document}
\centerline {\bf Noncommutative deformation and essence of a
             postulate of isotropy Iliusin}
\centerline {\it Trinh V.K.}
\centerline {\it Department of physics of polymer and crystal,
MGU}

\bigskip
\large

1. We shall consider non-uniform continuous environment.
 It is known, that under influence
limiting force this environment will pass in a plastic status.
To the present time many models of plastisity with different
 determining ratio. One of phenomenology approaches to
 these models is that from microstructural mechanism it is reasonable
to keep only main and to correct determining ratio
by the data of experiments. However, all experiments on plasticity show,
that in a plastic status the field of deformation and stress has strong
fluctuation. Besides the diagram $ \sigma=f (e) $ for a stress and
deformation
is similar to the diagram $P=f (V) $ for pressure and volume of liquid
[1],
and in [2-4]
the phase transition in a end of a crack was already observed. So, we
 can consider process of plastic deformation critical (phase
 transition), in which the structure is spontaneously formed [5]. Density
of
distributions of probability of transition to plasticity is the solution
of the equations Fokker - Planck. However, taking into account potential
character of
process of deformation, instead of the equation Fokker - Planck we can
 use the equation Schrodinger. It means, that process of plastic
 deformations can consider as process with supersymmetry. Here tensor of
deformation
 is considered as the parameter of order. In work [6] A.A. Iliusin
has presented on discussion of konseption of trajectory of deformation in
the space $E_5 $. Actually, the space $E_5 $ -it is layer in
 bundles  of deformation [5].
It means, that a field of deformation we can consider as a gauge field,
in which the generalized trajectory of deformation Iliusin is
a section of a bundles of deformation. With the help of the identification
of
 generalized trajectory of deformation with a knots we have shown
topological characters of process of deformations connected with
invariant Witten. Thus, we have received density of
distributions of probability of transition to plasticity [6] $$
Z = \int\exp[-kS_{cs}(e)]{\cal D}e, $$ here $k- $ it
factor of elasticity, $S_{cs}- $ action Chern - Simomse as
$$ S_{cs} = \int\Gamma_{cs}, $$ where $ \Gamma_{cs} $ - form
Chern - Simonse for variety of deformation.

2. The space of deformation $P = \{E_n(x, t) \} $, with most general
 points of view, is noncommutative. A. Connes in work
[7] has constructed noncommutative geometry, in which instead of
researches of character of point space it is considered
properties of a ring of function. Further we shall build model
plastic deformation within the framework of this geometry. It means, that
we should construct noncommutative bundles of deformation [8].
Two basic objects in geometry Connes are associative
algebra $A $ and universal algebra of the forms $ \Omega(A) $ over
 algebra $A $. Here space of deformation $P = \{E_n(x, t) \} $
plays a role of associative algebra. The following problem is
choice of universal algebra Connes. Because of that, each choice of
this algebra there corresponds the certain type of differential
calculations, in work [8] we have chosen $ \Omega(P) $ in quality of
algebras Hoof, using, thus, differential
calculation such as Woronowicz. It is known, that space of
deformation $P = \{E_n \} $ is associative algebra, in
 frameworks of which can be determined derivative and
the differential form. So, we choose universal algebra
Connes $ \Omega(P) $ as algebra of the differential forms over
the space of deformation $P $. Thus, differential
 calculation on $ \Omega_D(P) $ has a type similar to such in work
[9], according to which least differential algebra of
algebras of cocycle $C(Der(P), P) $, containing space $P $,
is $ \Omega_D(P) $, being algebra $C(Der(P), P) $, where
$Der(P) $ - algebra of derivative on $P $. In the space $P = \{E_n \} $
the basis $ \{1, E_1, E_2, \ldots, E_{n^2-1}, n+1\ldots 9 \} $ is defined
. We consider $P $ Hermite space with the following
commutative rules $$ E_k
E_l = \delta_{k} +C^m_{kl}-\frac{i}{2}C^m_{kl} E_l, ~ [E_k, E_l] =
-2iC^m_{kl}E_m
$$ Where
$ \delta_{kl} = \frac{1}{2}Tr(E_kE_l), ~S^m_{kl} =S^m_{lk}, ~C^m_{kl} =
-C^m_{lk}, ~C^m_{kl} $ -
constants of structure, $S^m_{kl} $ - invariant of group $sl(n) $ [10]. In
space $ \Omega^1_D(P) $ following basis is defined $ \theta^k, ~k\in
\{1,2,\ldots,n^2-1\} $, with
 properties $ \theta^k(\partial_l) = \delta^k_l{\bf
1}, ~ \theta^k\theta^l = -\theta^l\theta^k,
E_k\theta^ = \theta^lE_k $.
So, differential calculation in
$ \Omega^1_D(P) $ is to usual external differentials:
$$dE_k = -C^m_{kl} E_m\theta^l, ~
d\theta^k = -\frac{1}{2}C^m_{kl}\theta^l\theta^m,$$
$$
 ~d(\theta^k\theta^j) =
d\theta^k\theta^j-\theta^k d\theta^j, ~d^2=0 $$.

Let's recollect, that the bundles of deformation gives us section being
a field
of deformation
or generalized trajectory of deformation Iliusin [5]. In conformity with
noncommutative
 geometry Connes, vector bundle on noncommutative space of deformation
$P $  will be either left, or right module, in which a trivial bundle
of deformation
is the free module, and the nontrivial bundle represents
 factorization of the free module above space $P $. Actually it is
the cotangent bundle
$ \Omega^1_D(P) $.

Now we shall construct action for process of noncommutative plastic
deformation.
According to procedure Connes, first of all we choose Hermite module
$H $ above the space of deformation $P $. It means, that on Hermite
 $P$ -module
$H $ there is such Hermite structure $h $, that for all $ \alpha, \beta\in
H,
h(\alpha, \beta) \in P $ and $h(\alpha A, \beta B) =
A^{*}h(\alpha, \beta) B $ for all
$A, B\in P $. Further it is necessary to choose connection $ \nabla $,
 being transformation
$H $ on $H\otimes\Omega^1_D $, for which the folowing conditions
are carried out $ \otimes (\phi A) =
(\otimes\phi) A + \phi\otimes dA $ $dh(\phi, \psi) =
h(\nabla\phi, \psi) +h (\phi, \nabla\psi) $
for all $ \phi, \psi\in H, ~A\in P $.

Connection $ \nabla $ is extended to linear maps of a rank 1
$D_v:~~H\otimes
\Omega^1_D\to H\otimes\Omega^1_D $ so that $D_v(\phi\otimes\alpha) =
(\nabla\phi)\alpha +
\phi\otimes d\alpha $ for all $ \phi\in H, \alpha\in \Omega^1_D $. Thus,
the curvature on $ \nabla $ is determined as maps $ \nabla^2:
H\to H\otimes\Omega^2_D $,
$ \nabla^2(\phi B) = (\nabla^2\phi)B, \phi\in H, B\in P $.
Then action of process of noncommutative
 plastic deformation on connection $ \nabla $ will be submitted in the
form:
$$
S = {\Vert\nabla^2\Vert}^2 = < \varphi, \varphi >,
$$
where $ \varphi a $ -component connection on calibration with scalar
product
$$
< \varphi, \varphi > = -\int\varphi(\*\varphi) =Tr[\varphi(\*\varphi)].
$$
So, we have received distribution of probability of transition to
plasticity
$$
Z = \int\exp(-kS),
$$
and $k$ it is factor of elasticity of environment.
 Distribution $Z $ depends from
connection $ \nabla $, determined on any calibration. With the purpose of
reception of the obvious form $Z $
let's consider the following example. First of all we shall
choose calibration $¥$, being
the unit in Hermite space $H $. Then we shall choose connection as
$ \nabla e=e\otimes\alpha, $ where $ \alpha\in \Omega^1_D,
~ \alpha = -{\bar\alpha},
{\bar\alpha} $ - interface. The form $ \alpha $ is a complex matrix
in sense Muschelishvili.
However in 3-dimention  case, in the frameworks of symplectic geometry,
it will have
complex structure. Thus, each element $ \phi\in H $ looks like
$ \phi=eB $, where $B\in P $. Then by definition we shall receive
$ \nabla\phi = (\nabla e)B+ e\otimes dB $. In differential geometry $B $
and $ \alpha $
refer to as  elements $ \phi $ and $ \nabla $ on calibration $e $.
A component $ \varphi $ of the
curvature $ \nabla^2 $ on calibration $e $ is chosen as
$ \varphi=d\alpha + (\alpha)^2 $,
then in result we shall receive $ \nabla^2=e\otimes\varphi $.
If we shall present the form
$ \alpha $ in a usual complex kind $ \alpha=B_k\theta s^k+iE_l\theta^l,
 ~B, E\in P $,
that after action of the operator $d $ and $ \alpha $,
after calculation $ \alpha^2 $ and putting
results in expression for $\varphi $ we shall receive
$$
\varphi = \frac{1}{2}([B_k, B_l]-C^m_{kl}B_m)\theta^k\theta^l.
$$
So, the action has the form
$$
S = -\frac{1}{8}\sum\limits_{k,}Tr\{([B_k, B_l] -C^m_{kl}B_m)^2 \}
$$
It is necessary to notice, that this integral on Hermite module $H $ -
it is a $K$ - cycle Connes.
However, Connes has not defined the concept of " secondary calculation ",
therefore we shall calculate integral
$Z $ by a usual method.

3. In work [6] A.A. Iliusin has offered idea, according to which
determining
ratio for continuous initial isotropy  environments in area of rather
small deformations will be coordinated to a postulate isotropy:
it is  invariant
concerning orthogonal transformations in $E_5 $. In work [14,15] it is
described
determining ratio even  for final deformation. Nevertheless, within the
framework of model of
noncommutative  plastic deformation in a formalism of potential of
deformation
this postulate would become the theorem, which it is possible to prove.
Really in everyone
cotangent bundle always there is symplectic structure [13,8].
As shown above, cotangent space to space of deformation
$P = \{E_n (x, t) \} $
is the space $Der(P) $; cotangent space to space $P $
is $ \Omega^1_D(P) $. If Hamilton  vector field in a point $A\in P $
is  designated
through $H(A) $, the symplectic  structure will be the 2-form
$ \omega^2\in \Omega^2_D(P) $
so, that for all $A\in P $ the equation $ \omega^2(X, H(A))=XA $,
at all $X\in Der(P) $
has the unique solution   $H(A) \in Der(P) $. If $I $ is isomorfism  :
$ \Omega^1_D(P) \to Der(P) $, the Hamilton vector field will look like
$IdH $, and $H $ is function Hamilton  and the symplectic   2-form will be
as $ \omega^2=dH $. We designate the symplectic   variety as
$M^{2n^2}, \omega^2 $,
then for each vector field $H (A) $ there is an one-parametrical group
diffeomorphism  $g^t: M^{2n^2\to M^{2n^2}} $, which is named as
a phase flow
Hamilton :
$$
\frac{d}{dt} \mid_{t=0}g^t A=IdH(A)
$$
It is necessary to notice, that the orthogonal transformation is a special
case  of diffeomorphism .
So, it is possible to formulate a postulate of isotropy  in a formalism of
potential by the following
way . A determining ratio is invariant  concerning a flow Hamilton ,
or, as a special case, it is invariant  concerning orthogonal
transformations.
In frameworks  of commutative  geometry in work [13] V.I. Arnold  has
rendered,
 that the Hamilton
 phase flow keeps the symplectic  structure $ (g^t)^{\ *} \omega^2 =
\omega^2 $.
In ours noncommutative  case also it is possible to use a method Arnold
 with  only
one  difference, that the dimention of symplectic   space above $P $ is
equal $2n^2 $
 and points of
the spaces are matrixes. However, when the space $P $ satisfies still
to axioms of algebra Hoof,
it is impossible to use  above mentioned method because of unusual
differential
calculations such as Woronowicz . Further, by definition, we have $
\omega^2=
dH = \sum\frac{\partial H}{\partial e_i}de_i $ and this form is invariant
  concerning a
flow $g^t $.
It means, that in a condition of additivity the quantity
 $ \frac{\partial H}{\partial e_i} $
 is also invariant  concerning transformations $g^t $.

In work [11] in singular  model determining ratio is submitted as:
$$
d\sigma_i = \frac{\partial f_l}{\partial e_i}dk_l
$$
 where $f_l $  - the current limiting surface. Here  those  $dk_l $
are not zero,
for which on the appropriate surface $f_l $ the infringements are
performed,
 i.e are made
$ (\frac{\partial f_l}{\partial e_i})d\sigma\ge 0 $. Now, if we use
the concept  of
potential $H $ (free energy) [6,12], we shall receive
$$
\sigma_i = \frac{\partial f_l}{\partial e_i}
$$
In our model the plastic deformation is the fluctuation  process, in which
 the free energy plays a role of the function Hamilton of system [5].
 Really, determining
ratio in a formalism of potential is invariant  concerning orthogonal
transformation.

4. Thus, general  model of plastic deformation is following .
 The process of deformation Iliusin is given.
 For construction of the general  theory it is necessary,
first of all, to refuse
from traditional performance of the elastic theory of field,
 within the framework of which the static processes with commutative
 quantities  are described only.
The space of deformation should be considered  as noncommutative .
 Under action of any
forces this space will proceed  in cotangent bundle.
When the action will achieve
limiting value the status of deformation will be in strong fluctuation .
Then we
 receive distribution of probability of transition to plasticity.
Besides thanking
noncommutativity  we shall receive the symplectic  structure.
Symplectic  space,
constructed above noncommutative  space of deformation, it is more than
traditional space of the classical
 theories. Only in this case it is possible to understand the essence
of a postulate of isotropy Iliusin.
And, may be, all other results of the classical theory can be received as
special cases of this general  model. One of such results is the essence
of the
sectors Koiter , about what will be informed in following clause.


\begin{thebibliography}{99}
\bibitem{1} M.Ausloos: Solid Stat. Comm., 1986, vol 59,
No 6, p.401-404.
23,1986, p.2509-2512.
\bibitem{2} V.G.Gargin, Superfirm materials,
                1982, No.2, c.17-20.
\bibitem{3} V.A.Pesin, N.N.Tkachenko, L.I. Fedchuk, JFCh , 1979, Volume
53 No.2,
\bibitem{4} Li-Shing Li, R.J.Pabst, Materia Sci., 1980, vol 15, no 10, p
2861
\bibitem{5} Trinh V.K. : kan. diss., MGU , 1993.
\bibitem{6} A.A. Iliusin , Mechanics of continuous environment , MGU 1990.
\bibitem{7} Connes A., Noncommutative differential geometry, Publ.
I.H.E.S.62,257 (1986).
\bibitem{8} A.T. Fomenko,  Simplectis  geometry, MGU  1990.
\bibitem{9} Daboit-Violette M., Kerner R., Madore J., J.Math. Phys., 32
(2), 1990.
\bibitem{10} Madaran A.J., Comm. Math. Phys., 11,77 (1986).
\bibitem{11} V.D. Kiusnikov, Physical and mathematical bases of durability
and
     Plasticity, MGU 1994.
\bibitem{12} YU.N. Rabotnov, Mechanics of a deformable body, M.:Sciens
1988.
\bibitem{13} V.I. Arnold , Mathematical methods of the classical
mechanics, Œ.Sciens  1989
\bibitem{14} Tsakmakis Ch., Arch. Mech., 1997,49, no 4, c 677-695.
\bibitem{15} YU.N. Sevchenko, Appl. Mech.  (Kiev), 1999,35, no 1, c 14-27.
\end{thebibliography}
\end{document}